\documentclass[pdflatex,sn-mathphys,Numbered]{sn-jnl}
\usepackage{graphicx}%
\usepackage{multirow}%
\usepackage{amsmath,amssymb,amsfonts}%
\usepackage{amsthm}%
\usepackage{mathrsfs}%
\usepackage[title]{appendix}%
\usepackage{xcolor}%
\usepackage{textcomp}%
\usepackage{manyfoot}%
\usepackage{booktabs}%
\usepackage{algorithm}%
\usepackage{algorithmicx}%
\usepackage{algpseudocode}%
\usepackage{listings}%
\usepackage{wrapfig}
\usepackage{url,hyperref} 
\usepackage{color}
\usepackage{ulem}
\usepackage{verbatim}
\usepackage{float}
\usepackage{natbib}
\newcommand{\req}[1]{Eq.\,({\ref{#1}})}

\begin{document}
\title[Superheavy \& Ultradense Matter]{Superheavy Elements and Ultradense Matter}
\author{\fnm{Evan} \sur{LaForge}}\email{evanl@arizona.edu}
\author{\fnm{Will} \sur{Price}}\email{wprice@arizona.edu}
\author{\fnm{Johann} \sur{Rafelski}}\email{johannr@arizona.edu}
\affil{Department of Physics, The University of Arizona, Tucson, AZ 85721,USA}

\date{August 28, 2023,\href{https://doi.org/10.1140/epjp/s13360-023-04454-8}{Eur. Phys. J. Plus (2023) 138:812}}

\abstract{In order to characterize the mass density of superheavy elements, we solve numerically the relativistic Thomas-Fermi model of an atom. To obtain a range of mass densities for superheavy matter, this model is supplemented with an estimation of the number of electrons shared between individual atoms. Based on our computation, we expect that elements in the island of nuclear stability around $Z = 164$ will populate a mass density range of $36.0 - 68.4$ g/cm$^3$. We then extend our method to the study of macroscopic alpha particle nuclear matter condensate drops.}



\maketitle
\date{V2 of August 28, 2023, Accepted: 7 September 2023, published 15 September 2023\\\href{https://doi.org/10.1140/epjp/s13360-023-04454-8}{Eur. Phys. J. Plus (2023) 138:812}}

\section{Introduction}

\noindent
We explore properties of superheavy elements with nuclear charge $Z \geq 104$ in a simplified model approach. Our primary objective is to improve the systematic understanding of the mass density of such elements. Prior work relied on the Hartree-Fock method~\cite{Fricke1971}, which provides a detailed insight into the microscopic atomic structure. We instead use the relativistic Thomas-Fermi model (RTFM)~\cite{Muller:1974fh}, which allows us to significantly enlarge the range of atomic charge $Z$ and provide a more comprehensive view of superheavy elemental properties.

\subsection{Ultradense Matter}

\noindent
Our study of mass density is motivated by the possibility of Compact Ultradense Objects (CUDOs)~\cite{Rafelski:2013,Dietl:2012}, which we define here as objects, typically astronomical bodies, which have a mass density greater than that of Osmium (22.59 g/cm$^3$~\cite{Lide:2005}), the element with the largest experimentally measured mass density. In particular, some observed asteroids surpass this mass density threshold. Especially noteworthy is the asteroid 33 Polyhymnia. Based on mass and size measurements obtained from independent sources, its mass density is computed as \cite{CARRY201298}
\begin{equation} \label{Poly}
 \rho = 75.3 \pm 9.6 \, \text{g/cm$^3$} \, .
\end{equation} 
Since the mass density of asteroid 33 Polyhymnia is far greater than the maximum mass density of familiar atomic matter, it can be classified as a CUDO with an unknown composition.

In this paper we suggest and explore the notion that CUDO matter could be composed of high-$Z$ superheavy, and `condensed' $\alpha$-nuclei matter elements beyond the known periodic table. Islands of nuclear stability are thought to exist for high-$Z$ elements, particularly around $Z=164$~\cite{Grumann:1969}, now called `hyperheavy' nuclei. Study of nuclear stability generated by mean-field models with shell structure  continued~\cite{Bender:2003jk,Sobiczewski:2007}, however this method is biased toward generating relatively uniform nuclear matter density. It is  the deformed typically cigar shape of otherwise uniform nuclear charge distribution which provides the reduction of the Coulomb repulsion effect (however, see Ref.\,\cite{Nazarewicz:2002obt}). Another self-consistent Hartree-Fock method allows recognition of bubble-shaped hyper-heavy ({\it i.e.\/} $Z>120$) nuclei~\cite{Decharge:1999}, followed up by  continued study of geometric bubble structure~\cite{Decharge:2003yyt}, which has been explored up to $Z=240$. The in depth reconsideration of shell effects~\cite{Afanasjev:2018mgv,Agbemava:2019msn} has stimulated consideration of toroidal nuclei with a lower charge density~\cite{Agbemava:2020ryo} explored up to $Z=186$. These explorations all indicate the possibility of quasi-stable superheavy elements beyond the periodic table. How such extreme atomic nuclei could be produced, {\it e.g.\/} in the environment of a neutron star, is also being considered~\cite{Veselsky:2022}.

The mass density of any element is dependent on atomic mass and the spacing between atoms. The atomic mass is mostly localized in the atomic nucleus; the electrons and their binding energy have a negligible effect on the atomic mass. We note that as $Z$ increases, the number of nucleons increases at a faster than linear rate: light elements have roughly the same number of neutrons and protons, but heavy elements have roughly 1.5 times as many neutrons as protons. This increase originates in the need to reduce the mutual Coulomb repulsion between $Z$ protons in the atomic nucleus. 

Another factor impacting the mass density of elements is the space occupied by each atom. As $Z$ increases, the strength of the Coulomb force on the electrons, as well as the increased degeneracy in high angular momentum states, prevents the distance between atoms from growing nearly as fast as the mass.  Our main effort in this work is to develop an understanding of how the space occupied by atoms varies across a very large domain of $Z$. We find that the mass density range of superheavy elements on average would grow a bit faster than linear with $Z$.

Our results on mass density allow us to hypothesize that if superheavy elements are sufficiently stable, they could exist in the cores of dense asteroids like 33 Polyhymnia. We refer to Ref.\,\cite{CARRY201298} for a full listing of asteroid mass densities. There are many more entries with anomalous CUDO density values, albeit with larger margins of error and thus reduced statistical significance for individual objects. However, considered as a group, one could argue that there is strong evidence that the case of asteroid 33 Polyhymnia is not singular.

Beyond our consideration of high-$Z$ superheavy elements, we will also explore alpha matter, in which nuclear matter is composed of a gas of interacting alpha particles~\cite{Clark:2023sin}.

\subsection{Overview}

\noindent
The RTFM, which is discussed in section~\ref{Sec:Method}, approximates the quantum structure of an atom as a fixed nuclear charge distribution surrounded by a Fermi electron cloud. The ionic size is naturally defined at the radius at which the RTFM electron density vanishes. At this radius, the nuclear charge will not be entirely neutralized by the electron cloud, resulting in a slightly ionized atom (an ion) with a net positive charge. This ion can be thought of as an atomic ``core,'' with all remaining unscreened electrons appearing in the conduction band.

Defining the radius of such an atomic core as $R_{\mathrm{core}}$, we know that the potential outside the core takes the form of a Coulomb potential
\begin{equation} \label{gammadef}
 eV(r) = \frac{-\gamma\alpha\hbar c}{r},\; r \geq R_{\mathrm{core}} \, ,
\end{equation}
where $e$ is the charge of an electron, $V$ is the combined electric potential of the nuclear charge and electron cloud, $\alpha$ is the fine structure constant, and $\gamma$ is the the net charge of the core ion.

Based on experimentally measured mass densities, we show numerically that within a predefined class of metals, $\gamma$ scales approximately linearly with $Z$. We consider two classes of metals: typical metals, defined as having similar mass densities to other metals near them on the periodic table, and heavy metals, defined as having the highest mass density of any element in their respective period. It should be noted that most metals are at least as dense as typical metals and thus have a mass density in between that of typical and heavy metals with similar atomic number. There are some exceptions, such as Europium ($Z=63$) and Ytterbium ($Z=70$) \cite{Lide:2005}, which are less dense than typical metals, though these exceptions are rare.

\begin{figure}[t]
 \centering
 \includegraphics[width=0.780\textwidth]{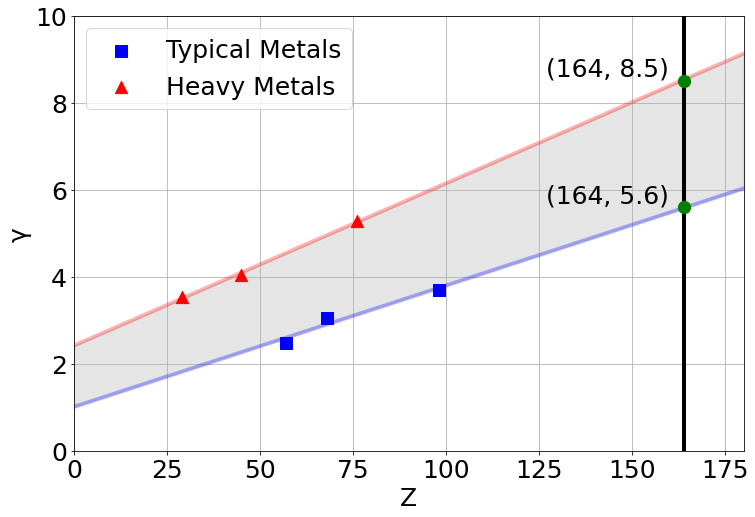}
 \caption{The number of electrons shared in the conduction band, $\gamma$, as function of the atomic number, $Z$, of a given element obtained according to the relativistic Thomas-Fermi model: nearly all metals will have a $\gamma$-value within the shaded area. At $Z = 164$, the intercepts with our linear extrapolation are marked to highlight the range of $\gamma$ for that value of $Z$.}
 \label{fig:Extrapolation}
\end{figure}

Figure~\ref{fig:Extrapolation} depicts the relationship between $\gamma$ and $Z$ for both considered classes of metals. For each value of $Z$ marked on the plot, the value of $\gamma$ was chosen by determining what value yielded the experimentally measured density when used as an input for the RTFM. We can then extrapolate $\gamma$ to higher $Z$ to be used as an input to the RTFM equations.

It should be noted that $\gamma$ was not restricted to integer values. This allows for $\gamma$ to become a fudge parameter, compensating for the systematic error, such as the use of spherical ions as a proxy for crystalline structure elements. In section~\ref{Sec:Dense}, we will use this extrapolation to determine the range of mass densities that superheavy elements are likely to fall within for a few values of $Z$, including $Z=164$.

We will also explore alpha matter using the RTFM in section~\ref{Sec:Alpha}. Unlike normal matter, significant screening of the Coulomb repulsion within a very large nucleus of alpha particles can occur by the electrons ``diving'' into the nucleus and potentially forming more alpha particles. This would allow for CUDO systems with much higher values of $Z$ to remain stable. We will use as an example $Z = 10^5$. The relative simplicity of the RTFM makes this value of $Z$ only slightly more computationally intensive than that of standard matter. However, using the more precise Hartree-Fock method would require calculating on the order of $10^5$ self-consistent orbitals, making it computationally challenging even today.
\section{The Relativistic Thomas-Fermi Model of the Atom} \label{Sec:Method}
\noindent
The RTFM atom consists of a fixed, spherically symmetric nuclear charge distribution surrounded by an electron cloud. The density of that cloud is determined by prescribing a relativistic Fermi energy $E_f$. The relativistic Fermi momentum is given by
\begin{equation} \label{eq:fermi_momentum}
 p_f^2(r)c^2 = [(E_f-eV(r))^2 - (m_e c^2)^2]\Theta(E_f-eV(r)-m_e c^2) \, ,
\end{equation}
where $\Theta$ is the Heaviside function. Note that at a certain radius, the right side of~\req{eq:fermi_momentum} vanishes. This defines the edge of the ion, $R_{\mathrm{core}}$, from the following condition 
\begin{equation}\label{Rcore}
 eV(R_{\mathrm{core}}) \equiv E_f - m_e c^2\;,\qquad E_f-m_ec^2 \equiv E_{f,\mathrm{NR}}\;,
\end{equation}
where $E_{f,\mathrm{NR}}$ is the non-relativistic Fermi energy.

The usual relation between the number density of electrons and the Fermi momentum also applies in the relativistic model as follows~\cite{Landau:1980}: 
\begin{align}\nonumber
 \rho_e(r) &=  \frac{p_f^3(r) c^3}{3\pi^2\hbar^3c^3}  \, ,\\ \label{eq:fermi_density}
 &=\frac{[(E_f-eV(r))^2 - (m_e c^2)^2]^{3/2}}{3\pi^2\hbar^3c^3} \, .
\end{align}
Note that writing $E_f$ in terms of $E_{f,\mathrm{NR}}$ in~\req{eq:fermi_density}, when $|eV|<< m_e c^2$ and $| E_{f,\mathrm{NR}} | << m_e c^2$, the conventional non-relativistic Thomas-Fermi model arises. 

The edge of the ion defined in~\req{Rcore} provides a natural method for calculation of mass density, which only requires determination of the self-consistent potential associated with the prescribed nuclear structure and Fermi energy. The size of the atom is obtained from the radius $R_\mathrm{core}$, while the mass will be sourced in the atomic nuclei.

The potential of the atom is determined by solving the Poisson equation for the electric potential $V(r)$ with the density of electrons in the Fermi surface~\req{eq:fermi_density} as the source term. In the following equation, the proton density $\rho_P(r)$ and the total density $\rho_T(r)$ are number densities, not charge densities. The Poisson equation then reads
\begin{equation} \label{DiffEQ}
 -\Delta V(r) = \frac{|e| \rho_T(r)}{\epsilon_0} = \frac{-e \rho_P(r) + e \rho_e(r)}{\epsilon_0} \, .
\end{equation}
The proton density $\rho_P(r)$ will be established in the following subsection.
\subsection{Nuclear Charge Distribution}
\noindent
In order to determine the potential, the nuclear charge density (i.e. proton density) must be specified. This was chosen to be a Fermi distribution normalized to have $Z$ charges, as is standard. This also allows for our computation to be directly compared to and validated by earlier studies which used the RTFM~\cite{Rafelski:1976ts}. The proton density is
\begin{align}\label{eq:FermiRho}
 \rho_P(r) &= \frac{C}{1+\text{exp}\left[\frac{4\log(3)(r-R_N)}{t}\right]} \\
 t &= 2.5 \,\mathrm{fm} \\
 \label{eq:Rn} R_N &= 1.2 A^{1/3}
\end{align}
where $t$ is the shell thickness, $R_N$ is the nuclear radius, and $C$ is a normalization constant defined by the integral 
\begin{equation} \label{NORM}
 \int_0^\infty 4 \pi r^2 \rho_P(r) \text{d}r = Z \, ,
\end{equation}
which is calculated numerically. The number of nucleons $A$ in the nucleus is assumed to be 
\begin{equation} \label{Nucleons}
 A = 2.5Z \, .
\end{equation}
This assumption provides a good approximation for superheavy elements. It will be altered when we consider alpha matter. With these assumptions in place, the nuclear distribution is entirely determined by the parameter $Z$. Note that the choice of $R_N$~\req{eq:Rn} fixes the density at the center of the nucleus to be roughly one nucleon for every $\sim7.2$ fm$^3$.

\subsection{Fermi Energy}
\noindent
The key RTFM parameter required is the Fermi energy $E_f$, which is defined as the energy up to which electron states are filled. The value of $E_f$ determines the ion-electron density. As determined by M\"uller and Rafelski~\cite{Muller:1974fh}, in the domain where the spontaneous pair production is energetically prohibited and stable atomic structure arises, the Fermi energy $E_f$ obeys the following inequality
\begin{equation}
 -m_e c^2 \leq E_f \leq m_e c^2 \, .
\end{equation}

The case $E_f = m_e c^2$ corresponds to a neutral atom while $E_f < m_e c^2$ corresponds to a cation. If the Fermi energy is greater than $m_e c^2$, then the binding energy must be positive, which is impossible. If the Fermi energy is less than $-m_e c^2$, then the binding energy of the least bound electron will be greater in magnitude than $2m_e c^2$, meaning there is enough energy in the system to produce electron-positron pairs; when pair production occurs, additional electrons will be spontaneously added to the system and their associated positrons will be ejected. 

We are most concerned with the densest of superheavy elements, so we will consider metals with a compact, crystalline structure. Within this structure, atoms are tightly packed together and share a few electrons in the conduction band. The nucleus, as well as the unshared electrons outside of the conduction band, can be thought of as an atomic core. Effectively, the core is an ion that is neutralized by the electrons in the conduction band.

We will approximate the volume of a single atom within a metal as a sphere with radius $R_{\mathrm{core}}$, which will enable us to calculate the mass density.

\subsection{Boundary Conditions}
\noindent
We will need to supplement the RTFM differential equation~\req{DiffEQ} with boundary conditions in order to specify a unique solution. In particular, we must know the potential at the center of the nucleus $V(r = 0)$ as well as its derivative. At the origin the electric field vanishes, so that the derivative of the potential is
\begin{equation}
 \frac{\partial V}{\partial r}\bigg |_{r=0} = 0 \, .
\end{equation}

The value of $V(0)$ must be determined in a way that is self-consistent with the two parameters $Z$ and $E_f$. In the absence of an analytic formula giving $V(0)$ as a function of these parameters, we will instead use the known asymptotic form of the potential for large $r$; outside the core, there is no charge so the solution takes the form of a Coulomb potential with the unscreened charge, $\gamma$, as shown in~\req{gammadef}. Recall that $\gamma$ refers to the net charge of the core, which is dependent on $Z$ and $E_f$. When considering a metal in a crystalline structure, $\gamma$ can be thought of as the number of electrons in the conduction band.

We use a shooting method~\cite{Press:2007} to numerically determine $V(0)$. There is a single value of $V(0)$ that makes the solution converge for large $r$, while all other values will result in a divergent solution. When the magnitude of $V(0)$ is too high, we find that the solution for $V$ will diverge above the expected $1/r$ form and approach infinity. When the magnitude of $V(0)$ is too low, the solution will change signs, which is impossible in a bound system. Thus, for any values of $E_f$ and $Z$, $V(0)$ can be found through bisection.

With this procedure in mind, we can determine the mass of the atom using $A\,m_N$, where $m_N$ is the mass of a bound nucleon and $A$ is the nucleon number, for simplicity we set $m_N\simeq m_p$ where $m_p$ is the proton mass. Other effects of nuclear and electron binding, such as the difference in mass between protons and neutrons the or the mass of electrons, have been neglected. The volume of the atom is obtained using $R_\mathrm{core}$ as its radius, which is dependent on both $Z$ and $E_f$. Because $\gamma$ is dependent on $Z$ and $E_f$, we can instead fix $Z$ and $\gamma$ as our two parameters. This allows the calculation of the mass density of superheavy elements based solely on the nuclear charge and the number of shared electrons in the conduction band.

\subsection{Applying the Method}
\begin{figure}[b]
 \centering
 \includegraphics[width=0.780\textwidth]{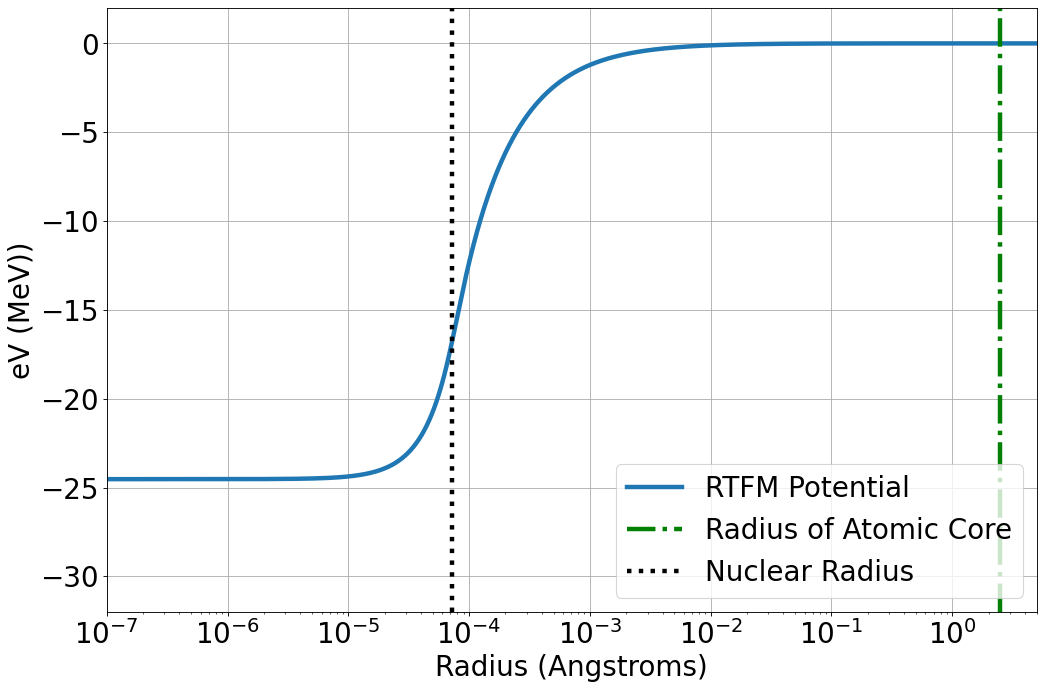}
 \caption{The potential energy of an electron, $eV$, as a function of the distance from the origin, $r$, for an RTFM atomic cation with $Z = 86$ and $\gamma = 2$. The nuclear and RTFM core radii have been marked with vertical lines.}
 \label{fig:Radon1}
\end{figure}

\noindent
We developed a C-program that solves the relevant differential equation~\req{DiffEQ} using a fifth order predictor corrector method for prescribed values of the parameters $Z$ and $\gamma$. In order to test the method, we considered in Figure~\ref{fig:Radon1} the results of this method applied to Radon ($Z = 86$), which were obtained using $\gamma = 2$. Radon was chosen because it is a noble gas with relatively high $Z$, so the Fermi surface electron cloud approximation of the RTFM is reasonable.

\begin{figure}[t]
 \centering
 \includegraphics[width=0.780\textwidth]{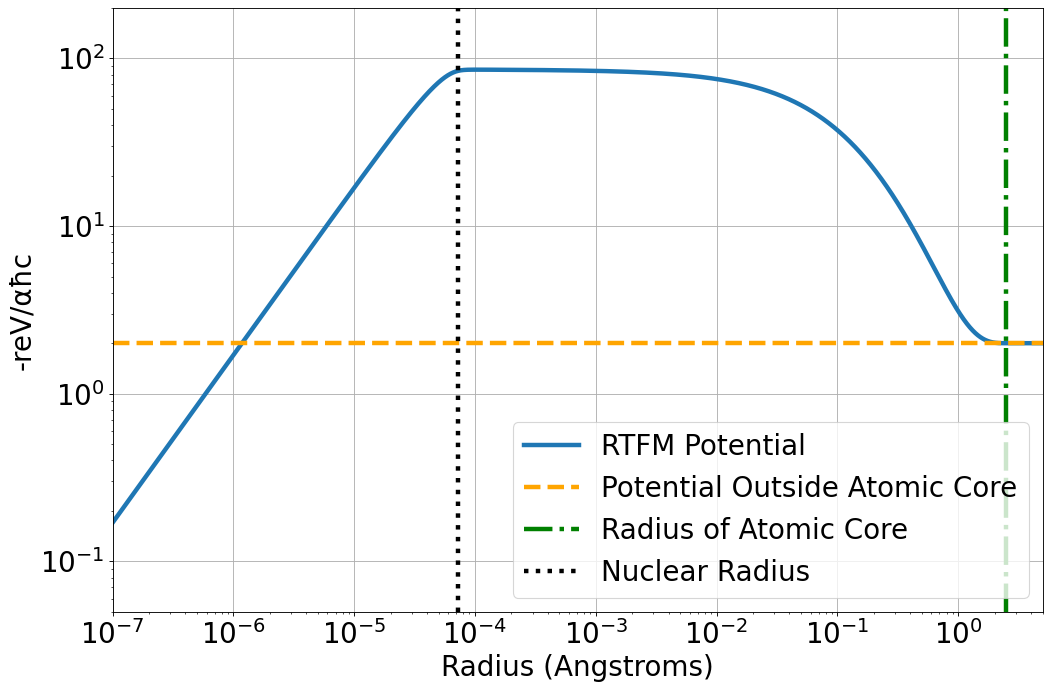}
 \caption{This figure depicts the same data as Figure~\ref{fig:Radon1}, with the y-axis non-dimensionalized to $-reV/\alpha\hbar c$ and log-scaled. The asymptotic potential outside the atomic core is also depicted and appears as a flat line at $+\gamma$.}
 \label{fig:Radon2}
\end{figure}

To visualize the asymptotic behavior of the solution, Figure~\ref{fig:Radon2} depicts the same numerical solution with $-\frac{reV}{\alpha \hbar c}$ instead plotted on the vertical axis. All values were made positive to enable log scaling of the vertical axis. In order to illustrate the convergence of this solution at the edge of the atomic core, the graph depicts the potential outside the atomic core $eV(r) = \frac{-\gamma \alpha \hbar c}{r}$, which, due to the scaling, appears as a flat line at $+\gamma$.

To verify the accuracy of the numerical method, we fed the RTFM potential into a numerical Schroedinger equation solver to determine if the potential for Radon ($Z = 86$) yielded the expected behavior. More specifically, we verified that the various electron shells were filled in the proper order according to~\cite{NIST_ASD}.

Additionally, we verified that there was a proportionately large energy gap between the least bound filled state and the most bound unfilled state, corresponding to high chemical stability. The least bound filled state was the 6p state with a binding energy of $7-8$ eV. The most bound unfilled state was the 7s state with a binding energy of $3-4$ eV. This is roughly half of the binding energy of the least bound filled state.

The experimentally measured value of the lowest ionization energy is 10.7 eV~\cite{NIST_ASD}. This is larger than the numerical result by about $25$\%, indicating the relative accuracy of the RTFM to describe properties of the noble gas radon: the qualitative features of the atomic structure is properly described by the self consistent RTFM potential.

\section{Mass density of superheavy elements}\label{Sec:Dense}

\noindent
Given $Z$, we will determine what unique value of $\gamma$ yields the correct mass density for a few elements with similar chemical properties. From these values of $\gamma$, we will linearly extrapolate a range of values of $\gamma$ for metals with a given value of $Z$. More specifically, we will consider typical metals and heavy metals. Typical metals are defined by having similar mass density to elements with similar atomic number, while heavy metals are defined as being the densest elements of their respective periods within the periodic table.

\begin{figure}[t]
 \centering
 \includegraphics[width=0.780\textwidth]{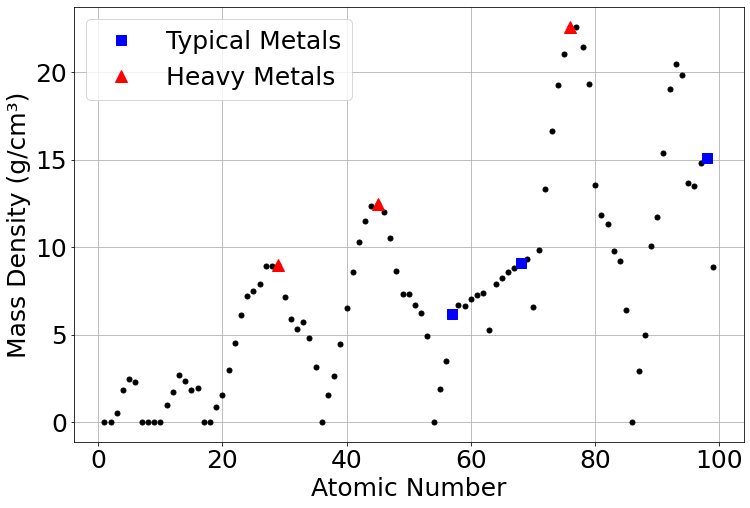}
 \caption{This figure depicts the density in g/cm$^3$ for all elements up through $Z = 99$, the limit of what has been experimentally measured.\cite{Lide:2005} Labeled on the graph are the typical metals, marked with blue squares, and the heavy metals, marked with red triangles, that were used for finding the baseline $\gamma$ values. The density values for all gasses were recorded at standard temperature and pressure.}
 \label{fig:Density Graphs}
\end{figure}

For typical metals, we chose Lanthanum ($Z = 57$), Erbium ($Z = 68$), and Californium ($Z = 98$). For heavy metals, we chose Copper ($Z = 29$), Rhodium ($Z = 45$), and Osmium ($Z = 76$). All of these elements are marked on Figure~\ref{fig:Density Graphs}. Note the humps in the graph produced by the varying chemical properties throughout individual periods on the periodic table.

With the elements of known mass density chosen, the next step is determining the appropriate values of $\gamma$ that yield these mass densities from the RTFM. This was done via bisection similar to the determination of $V(0)$. For each value of $Z$, a test value of $\gamma$ was chosen. For each test value, the potential was calculated and then used to determine $R_{\mathrm{core}}$. The mass density was then calculated using the equation below
\begin{equation} \label{eq:density}
 \rho_\mathrm{mass} = \frac{A m_p}{\frac{4}{3}\pi R_{\mathrm{core}}^3} \, ,
\end{equation}
where $m_p = 1.67 \times 10^{-24}$ g is the mass of the proton. This assumes that the volume occupied by one atom is equal to a sphere with the size of $R_\mathrm{core}$. Deviations from this assumption are compensated for by allowing $\gamma$ to take non-integer values.

Once the mass density was calculated, the test value of $\gamma$ was increased or decreased in order to better match the mass density until precision up to ~1\% was achieved. Table~\ref{tab:Gammas} and Figure~\ref{fig:Extrapolation} depict the values of $\gamma$ for all six considered input elements used in the linear extrapolation of $\gamma$ up to $Z = 180$. $\gamma$ ranges roughly from $2$ to $9$ in the range of $Z$ for which the RTFM is valid ($Z \gtrsim 25$).

\begin{table}[b]
 \centering
 \begin{tabular}{c|c|c|c}
 $Z$ & 57 & 68 & 98 \\
 \hline
 $\rho_\mathrm{mass}$(g/cm$^3$) & 6.15 & 9.07 & 15.1\\
 \hline
 $\gamma$ & 2.48 & 3.06 & 3.70 \\
 \end{tabular}
 
 \vspace{2em}
 
 \begin{tabular}{c|c|c|c}
 $Z$ & 29 & 45 & 76 \\
 \hline
 $\rho_\mathrm{mass}$(g/cm$^3$) & 8.96 & 12.4 & 22.59\\
 \hline
 $\gamma$& 3.54 & 4.04 & 5.28\\
 \end{tabular}
 \caption{For three input elements with value of $Z$ we show: The experimental mass densities~\cite{Lide:2005} $\rho_\mathrm{mass}$(g/cm$^3$) and our calculated $\gamma$-values. The top table section corresponds to typical metals, and the bottom table section corresponds to the heavy metals, see text.}
 \label{tab:Gammas}
\end{table}

Using the extrapolation of $\gamma$, we determine the range of $\gamma$ values that should appear near $Z=164$. For each $\gamma$ value, the RTFM is used to obtain the radius of the atomic cores, which enables calculation of the mass density of the metal. Because mass density increases monotonically with $\gamma$ (because the core size decreases as $\gamma$ increases), the actual mass density of superheavy element $Z=164$ is found between the mass densities calculated using these two extrapolated $\gamma$ values.

The extrapolations for typical and heavy metals respectively yield the equations
\begin{align}
 \gamma_\mathrm{typical} &= 0.0279 Z + 1.01 \\
 \gamma_\mathrm{heavy} &= 0.0373 Z + 2.42
\end{align}
as shown in Figure~\ref{fig:Extrapolation}. We are encouraged that for the lower range of $Z$ values for which the RTFM is valid ($Z \gtrsim 25$), $1.7 < \gamma < 3.4$, which is a reasonable range for the number of electrons in the conduction band for these metals. 

\begin{figure}[t]
 \centering
 \includegraphics[width=0.780\textwidth]{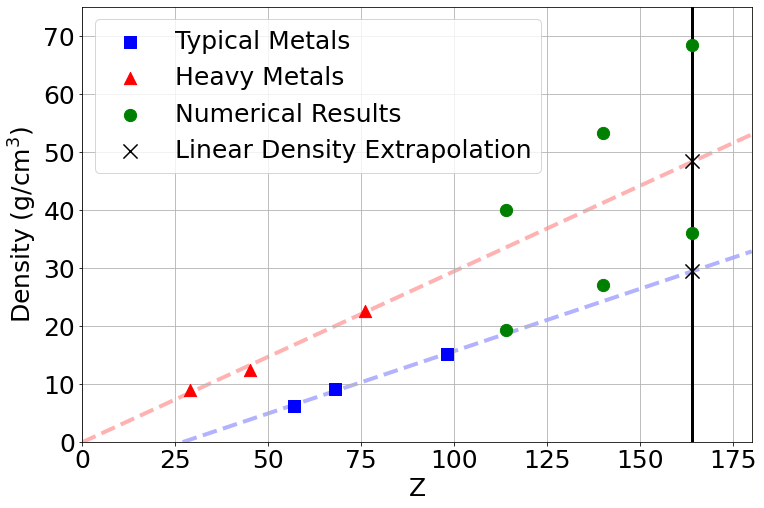}
 \caption{The predicted mass density bounds for superheavy elements in domains of atomic number $Z=$ 114, 140, and 164 (round (green) dots; input (triangles (red) and squares (blue); shown as function of atomic number $Z$. The vertical line indicates the predicted `island of stability' at/near to $Z=164$; the region $Z=114$ is already discovered. It is unlikely that $Z=140$ yields a stable nucleus, but this value of $Z$ makes the shape of our result more clear compared to dashed lines presenting a naive linear extrapolation.\label{fig:DenseExtra}}
\end{figure}

We see that $\gamma_\mathrm{typical} = 5.6$ and $\gamma_\mathrm{heavy} = 8.5$ for $Z = 164$. The RTFM solution then gives us the typical and heavy metal core radii $R_{\mathrm{core}} = 1.656, \, 1.338$ \AA, respectively. From~\req{eq:density} we then predict that $Z=164$ will have a density between $36.0$ and $68.4$ g/cm$^3$, see Figure~\ref{fig:DenseExtra}. These predictions align with results of Fricke~\cite{Fricke1971}, who predicted that $Z=164$ would have a density of $46$ g/cm$^3$.

Note that our predicted mass density range will work not only for $Z = 164$ but for elements with similar atomic number as well. This is because increasing or decreasing $Z$ by small amounts will make very little difference on the mass density range. Thus, all elements near $Z = 164$ will have similar mass density ranges, and if there is a broad enough island of nuclear stability, then atoms will likely populate the range of mass densities from $36.0$ to $68.4$ g/cm$^3$.

We applied our method to determine a rough mass density range for any metals with $Z$ between $110$ and $118$, which elements have been discovered, though their mass densities are yet to be experimentally measured. A prior study of elements ranging from $Z=103$ to $Z=111$ using density functional methods predicts a range of mass densities of $14.4-27.3$ g/cm$^3$~\cite{Gyanchandani:2011}. The mass densities of elements $113$ to $116$ were predicted by Fricke to be between $13.6$ and $15.1$ g/cm$^3$~\cite{Fricke1971}.

Because $Z = 114$ is the midpoint of the range between $Z = 110$ and $Z = 118$, these elements should have mass densities within the bounds obtained for element $Z = 114$. Our extrapolation yields the range $4.2 < \gamma < 6.7$ for $Z = 114$. From these values, the lower bound on mass density is $19.3$ g/cm$^3$, and the upper bound is $39.9$ g/cm$^3$, see Figure~\ref{fig:DenseExtra}. Our result is near to the prior predictions quoted above but allows a slightly larger upper-range. It is clear that these relatively recently discovered elements could have surprisingly high mass density. However, they neither have a high enough mass density to explain CUDOs like asteroid-33 Polyhymnia, nor the stability that is required. 

It is worth considering whether we could consider a naive linear fit of mass density and avoid numerical calculations, such as the dashed lines in Figure~\ref{fig:DenseExtra}. However, because mass density is the product of both nuclear and atomic properties, it is unlikely that its form will be as simple as a linear $Z$ dependence originating primarily in the change of nuclear mass. We believe that an understanding of the number of shared electrons, $\gamma$, and evaluation of the cationic size which incorporates an understanding of atomic structure is necessary to obtain mass density of unknown superheavy elements. Thus, our extrapolation of $\gamma$, which considers nuclear and atomic structure is likely to be much more reliable in its description of superheavy elements.

To increase credibility of this argument, we study our $\gamma$ extrapolation predictive power pertaining to the properties of light elements. Using our linear form of $\gamma$, light elements are predicted to share between 1 and 3 electrons in the conduction band, in agreement with known elemental properties in domain $Z=13$ (Aluminum). In addition, light metals are predicted to have reasonable density ranges. For example, Aluminum has a predicted density range of 1.36 g/cm$^3$ to 6.02 g/cm$^3$. Correcting for our model's assumption of nuclear weight $A = 2.5*Z$, the actual density range should be about $20\%$ lower. Thus, we predict that elements near $Z = 13$ should fall within the range 1.1  g/cm$^3$ to 4.9 g/cm$^3$, which contains in the middle of this range the actual Aluminum density of 2.7 g/cm$^3$ \cite{Lide:2005}. As seen in Figure~\ref{fig:Density Graphs}, there are many elements with densities near to the typical metals but very few elements with densities near to the upper heavy metal limit. Thus, it is expected that most metals near $Z = 13$ have densities closer to 1.1 g/cm$^3$ than to 4.9 g/cm$^3$, which reflects reality. Thus our linear in $\gamma$ extrapolation method results in an appropriate density range for light metals, even though this approach has been tuned to model heavy rather than light elements.

If instead of using our atomic structure method one tried to predict the density of light elements using a linear fit of mass density as seen in Figure~\ref{fig:DenseExtra}, this would predict that near to $Z=13$ the density range between -3.07 g/cm$^3$ and 3.75 g/cm$^3$. As mentioned before, most elements near $Z = 13$ should have densities near the lower bound. However, this results in non-physical negative density values.

\section{Alpha matter}\label{Sec:Alpha}

\noindent
Much of the work in the previous section was built on the assumption of nuclear stability of superheavy elements and there is obvious interest in $Z>>164$ to obtain CUDOs of greater mass density. However, given that the Coulomb energy increases rapidly with $Z$, islands of nuclear stability are unlikely to occur beyond $Z=164$, and it is unclear whether $Z=164$ will be stable enough to explain the mass density of asteroid 33 Polyhymnia. However, a different high-density nuclear structure could lead to stable ultradense matter.

As an example, we explore alpha matter: nuclear matter composed of alpha particles in a Bose-Einstein condensate-like configuration. This configuration could be much less tightly packed than standard nuclear matter. For a recent review of this nuclear structure, see Clark and Krotscheck~\cite{Clark:2023sin}. For large enough $Z$, the proton density can be much lower while still maintaining nuclear attraction while the electron cloud is able to nearly overlap with the proton distribution to roughly counteract the Coulomb force responsible for nuclear instability. While this section focuses on condensates of alpha particles for simplicity and conciseness, similar matter could form using carbon-12 nuclei, oxygen-16 nuclei, or any other particularly stable bosonic nuclei.

To model the large, loosely packed cloud of alphas, we will use the Gaussian charge distribution
\begin{equation}
 \rho_P(r) = \rho_P(0) e^\frac{-r^2}{2R_N^2}\;,
\end{equation}
where $R_N$ is some arbitrary nuclear radius and $\rho_P (0)$ is defined from the constraint~\req{NORM}. Unlike Fermi charge distribution shape \req{eq:FermiRho} the Gaussian nuclear distribution can be integrated analytically to determine $\rho_P(0)$.
\begin{equation}
 \rho_P(0)=\frac{Z}{(2\pi)^{3/2} R_N^3}
\end{equation}
Note that $R_N$ is more of a characteristic size than an actual approximation of the size of the nucleus. A better description of the nuclear charge range is about $3R_N$, because about 97\% of the nuclear charge is contained within that radius. 

As example we consider case $Z=10^5$. We will choose $\gamma = 250$ because this large value of $Z = 10^5$ is well beyond the range of the extrapolation from the previous section. The ratio of $\gamma$ to $Z$ is roughly one order of magnitude smaller than the ratios for previous elements. Thus, $\gamma = 250$ is an underestimate for the true value of $\gamma$. As $\gamma$ increases, the volume of the core decreases, hence we believe that our assumption about $\gamma$ should provide a reasonable lower bound for the mass density of these forms of alpha matter. Because $\gamma$ is an underestimate, and density increases with $\gamma$, the predicted mass values for various forms of alpha matter should also be seen as underestimates.

To explore the atomic properties of this form of nuclear matter broadly, we will consider a wide variety of characteristic nuclear radii $R_N$. In Table~\ref{tab:alphadense} the considered characteristic radii are depicted and central charge density (left two columns) is obtained for $Z=10^5$, assuming $\gamma = 250$. Table~\ref{tab:alphadense} further lists the central potential energy $eV(0)$, the Fermi momentum at the nuclear radius $p_f{R_N}$, and the mass density $\rho_\mathrm{m}$ of the corresponding alpha matter for the considered values of $R_N$. We note that at the smallest nuclear radius, given that the potential at the origin exceeds twice the muon mass $2m_{\mu} = 212$ MeV/$c^2$, an additional charge distribution must be considered to accommodate for a few muons. However, this effect is not significant here. 

\begin{table}
 \centering
 \begin{tabular}{c|c|c|c|c}
 $R_N$[fm]& $\rho_P(0)$[fm$^{-3}$] & $eV(0)$[MeV] & $p_f|_{R_N}$[MeV/c] & $\rho_\mathrm{m}$[g/cm$^3$] \\
 \hline
 50 & $5.08\times10^{-2}$& -218 & 183 & 2240000\\
 100 & $6.35\times10^{-3}$& -109 & 91.6 & 1820000\\
 1000 & $6.35\times10^{-6}$& -10.4 & 9.17 & 712000\\
 10000 & $6.35\times10^{-9}$& -0.702 & 0.924 & 101000\\
 20000 & $7.94\times10^{-10}$& -0.244 & 0.466 & 36000\\
 50000 & $5.08\times10^{-11}$& -0.0480 & 0.189 & 5350
 \end{tabular}
 \caption{For $Z=10^5$ and chosen characteristic Gaussian nuclear radius $R_N$ we show from left to right: potential energy at the origin $eV(0)$, Fermi momentum at the nuclear radius $p_f|_{R_N}$, and mass density $\rho_\mathrm{m}$.}
 \label{tab:alphadense}
\end{table}

We note in Table~\ref{tab:alphadense} that between the three largest radii $R_N$, the Fermi momentum dips below 0.511 MeV/$c$, marking the onset of the transition to a non-relativistic electron distribution: In the non-relativistic limit, every alpha has two electrons bound to it, forming an atomic condensate that will not be considered in this work. However, in systems where the alpha particles are nearer to standard nuclear density, the electrons can be shared across all of the alphas, which we do consider in this study.

It should be noted that the nuclear radius of 50 fm (densest option listed in Table~\ref{tab:alphadense}) still has a $\sim 30\%$ lower nuclear density at the origin than standard nuclear matter. For $Z=10^5$ the size of the nuclear distribution $3R_N = 150$ fm is less than half of the Compton wavelength of an electron. Therefore, the electron cloud is not capable of fully counteracting the nuclear charge density and cannot fully eliminate the Coulomb repulsion between alpha particles; our numerical results indicate that only $90\%$ of the charge could be screened near the origin. Therefore the resulting potential well near the origin is fairly deep.

\begin{figure}
 \centering
 \includegraphics[width=0.780\textwidth]{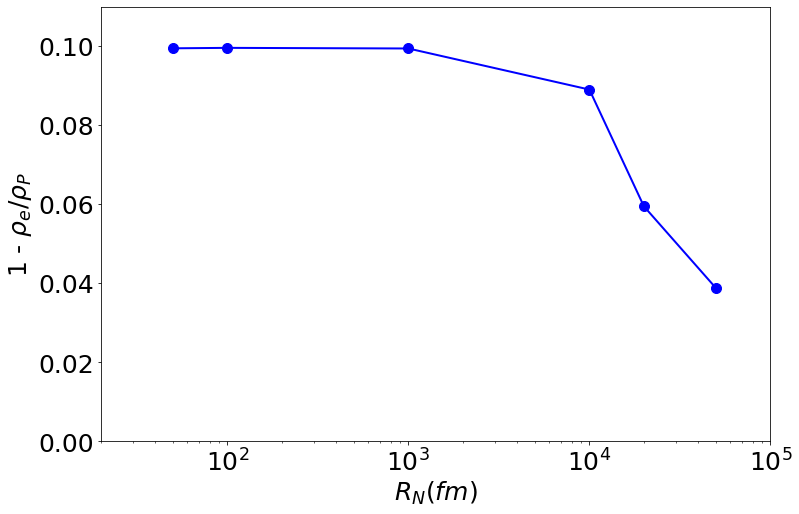}
 \caption{The proportion of nuclear charge at the origin not neutralized by electrons, defined as 1 - $\rho_e/\rho_P$, as a function of alpha-matter Gaussian size $R_N$.}
 \label{fig:Cancel}
\end{figure}

Figure~\ref{fig:Cancel} characterizes this situation more fully: We show the proportion of proton density at the origin that is not neutralized by electrons. For smaller radii, the proportion of charge not neutralized by the electron distribution is roughly 10\%, which still corresponds to a significant amount of cancellation. For higher radii, this proportion decreases significantly as the two distributions overlap nearly perfectly. This results in the Coulomb repulsion decreasing significantly as well, indicating the importance of further studying the nuclear alpha matter's stability with large $Z$.

\section{Discussion}\label{sec:Future}

The purpose of this study was to determine whether CUDOs with extreme mass density could be achieved without the need for the usually invoked strange or dark matter. We have done this while exploring two different nuclear systems using the relativistic Thomas-Fermi model. From the exploration of both standard nuclei and alpha matter, it is clear that both types of nuclear matter could explain the density seen in CUDOs such as asteroid-33 Polyhymnia.

Considering elements between $Z=110$ and $Z=118$, we predict that their mass densities are less than $40$ g/cm$^3$. It is therefore unlikely that these elements have a large enough mass density to explain the extraordinary mass density of asteroid-33 Polyhymnia; moreover, these elements are already known to lack the required stability. However, elements in the other theoretical island of nuclear stability near $Z = 164$, which we predict to populate mass density values between $36.0$ and $68.4$ g/cm$^3$, are reasonable candidates. If some significant part of the asteroid were made of these superheavy metals, it is plausible that the higher mass density could be near the experimentally measured value listed in~\req{Poly}. If there are other superheavy elements well beyond $Z = 164$, these could have an even higher mass density, further increasing the chances that our stated hypothesis of CUDOs containing superheavy elemental matter is correct.

We also applied the RTFM to alpha matter, which is another nuclear form of matter capable of explaining the large mass density observed in CUDOs. The forms of alpha matter described in Table~\ref{tab:alphadense} for $Z=10^5$ have mass densities of thousands or even millions of g/cm$^3$. This means that even a relatively small amount of alpha matter present in CUDOs like 33 Polyhymnia could result in the large observed mass density. These results, while preliminary, show the wide array of exploration that can be done using the computational efficiency of the RTFM. in such large alpha-nuclear matter drops there is nearly full cancellations of nuclear Coulomb repulsion.

There are a number of ways in which this work could be expanded in the future. The nuclear distribution was kept fixed, though electrons intersecting the nucleus would have likely resulted in alterations to the nuclear distribution and are capable of enhancing the stability of high $Z$ nuclei. Calculation of the nuclear distribution in parallel with the electron distribution would be required for this improvement, which could be accessible in the realm of the Walecka model of high-$Z$ nuclei~\cite{Walecka:1974qa}, or using and extrapolating methods addressing the alpha-matter nuclei~\cite{Clark:2023sin}. 

Joint computation of atomic and nuclear structure is particularly needed for alpha matter due to the significant overlap between the nuclear charge distribution and the electron cloud. In this instance a self-consistent Bose-Einstein condensed shell configuration could arise~\cite{Sun:2018pra,Carollo:2022Nat}. In addition, the electron mass in this model was assumed to be fixed, but using an effective mass would make more sense when such large binding energies are being considered. The exploration of atomic structure of alpha matter could also be more in depth, with more detailed models developed to describe their properties with more precision than the calculations made in this paper.

Our study has demonstrated that the relativistic Thomas-Fermi model of the atom is a useful tool. Because of its broad applicability and relatively small computation time, in particular for high $Z$, it should not be overlooked in applications to the atomic properties of CUDO matter.\\[0.4cm]

\noindent
{\bf Data Availability Statement:} No Data associated in the manuscript

\end{document}